\documentstyle[aps,pre,twocolumn,psfig,floats]{revtex}
\begin{document}
\wideabs{
\title{Force Distribution in a Granular Medium}
%\draft
\author{Daniel M. Mueth, Heinrich M. Jaeger, Sidney R. Nagel}
\address{The James Franck Institute and Department of Physics\\
	The University of Chicago\\
	5640 S. Ellis Ave., Chicago, IL 60637}
\maketitle

\bibliographystyle{prsty}

\begin{abstract}
We report on systematic measurements of the distribution of normal forces exerted by granular material under uniaxial compression onto the interior surfaces of a confining vessel.  
Our experiments on three-dimensional, random packings of monodisperse glass beads show that this distribution is nearly uniform for forces below the mean force and decays exponentially for forces greater than the mean.  
The shape of the distribution and the value of the exponential decay constant are unaffected by changes in the system preparation history or in the boundary conditions.  
An empirical functional form for the distribution is proposed that provides an excellent fit over the whole force range measured and is also consistent with recent computer simulation data.
\end{abstract}
\pacs{46.10.+z,05.40.+j,83.70.Fn,81.05.Rm}
%46.10.+z Mechanics of Discrete Systems 
%05.40.+j Fluctuation phenomena, random processes, and Brownian motion
%83.70.Fn Granular solids
%81.05.Rm Porous materials; granular materials 
}

\section*{Introduction}
Granular materials have a rich set of unusual behavior which prevents them from being simply categorized as either solids or fluids \cite{jaeger96b}.  
Even the most simple granular system, a static assembly of noncohesive, spherical particles in contact, holds a number of surprises.  
Particles within this system are under stress, supporting the weight of the material above them in addition to any applied load.  
The inter-particle contact forces crucially determine the bulk properties of the assembly, from its load-bearing capability \cite{travers87,guyon90} to sound transmission \cite{leibig94,liu92,melin94} or shock propagation \cite{potapov96d,sinkovits95}.  
Only in a crystal of identical, perfect spheres is there uniform load-sharing between particles.  
In any real material the slightest amount of disorder, due to variations in the particle sizes as well as imperfections in their packing arrangement, is amplified by the inherently nonlinear nature of inter-particle friction forces and the particles' nearly hard-sphere interaction.  
As a result, stresses are transmitted through the material along ``force chains'' that make up a ramified network of particle contacts and involve only a fraction of all particles \cite{ammi87,liu95,baxter97}.  

Force chains and spatially inhomogeneous stress distributions are characteristic of granular materials.  
A number of experiments on 2D and 3D compression cells have imaged force chains by exploiting stress-induced birefringence \cite{ammi87,liu95,baxter97,dantu57,dantu67,wakabayashi59,travers88,delyon90,dufresne94,howell97}.  
While these experiments have given qualitative information about the spatial arrangement of the stress paths inside the granular assembly, the quantitative determination of contact forces in three dimensional bead packs is difficult with this method.  
Along the confining walls of the assembly, however, individual force values from all contacting particles can be obtained.  
Liu et al.'s experiments \cite{liu95} showed that the spatial probability distribution, $P(F)$, for finding a normal force of magnitude $F$ against a wall decays exponentially for forces larger than the mean, $\overline{F}$.  
This result is remarkable because, compared to a Gaussian distribution,  it implies a significantly higher probability of finding large force values $F \gg \overline{F}$.

A number of fundamental questions remain, however. 
While several model calculations \cite{liu95,coppersmith96}, computer simulations \cite{radjai96b,radjai97c,radjai97b,luding97,thornton97} as well as experiments on shear cells \cite{miller96} and 2D arrays of rods \cite{baxter97} have corroborated the exponential tail for $P(F)$ in the limit of large $F$, other functional forms so far have not been ruled out \cite{eloy97}.  
Furthermore, there has been no consensus with regard to the shape of the distribution for forces smaller than the mean.  
The original ``q-model'' by Coppersmith et al. \cite{coppersmith96} and Liu et al. \cite{liu95} predicted power law behavior with $P(F) \propto F^{\alpha}$ and $\alpha \approx 2$ for small $F$, while recent simulations by Radjai et al. \cite{radjai96b,radjai97c,radjai97b} and Luding\cite{luding97} found $\alpha \le 0$.  
So far, experiments have lacked the range or sensitivity required for a firm conclusion.  
The roles of packing structure and history, identified in much recent work as important factors in determining stresses in granular media, have not yet been explored experimentally in this system.  
Finally, the existence of correlations between forces remains unclear.  
Shear cell data by Miller et al. \cite{miller96} have been interpreted as an indication for correlations between forces against the cell bottom surface. 

In this paper we present results from a set of systematic experiments designed to address these issues.  
We have refined the carbon paper method \cite{liu95,delyon90,dufresne94} for determining the force of each bead against the constraining surface and are now able to measure force values accurately over two orders of magnitude.  
With this improvement we are able to ascertain the existence of the exponential behavior and to obtain close bounds on its decay constant in the regime $F > \overline{F}$.  
For $F < \overline{F}$ we find that $P(F)$ flattens out and approaches a constant value.   
In addition, our experiments investigated the effects of the packing history. 
We also studied both the influence of the boundary conditions posed by the vertical container walls on the distributions of forces $P(F)$ as well as the spatial correlations in the arrangement of beads due to crystallization near a wall during system preparation.  
None of these variations on the experiment are found to influence $P(F)$ significantly.  
Finally, we have also measured the lateral correlations between forces on different beads and find that no correlations exist.

\section*{Experimental Method}
\begin{figure}
\centerline{
\psfig{file=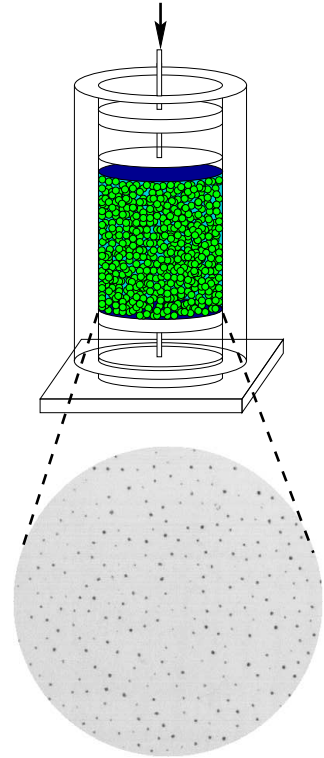,width=1.6in}}
\vspace{2ex}
\caption{
Sketch of the apparatus used for experiments with ``floating walls.''  
The lower piston is fixed and the cylinder is supported by friction with the bead pack.  
A load is applied to the upper piston and the beads press the carbon paper into white paper, leaving marks which are used to determine the contact forces.  
A detail of the obtained raw data is shown in the photograph (field of view:  76 mm across).
}
\label{fig:apparatus}
\end{figure}

The granular medium studied was a disordered 3D pack of 55,000 soda lime glass spheres with diameter $d =  3.5\pm0.2$~mm. 
The beads were confined in an acrylic cylinder of 140~mm inner diameter.  
The top and bottom surfaces were provided by close-fitting pistons made from 2.5~cm thick acrylic disks rigidly fixed to steel rods.  
The height of the bead pack could be varied, but experiments described in this paper were performed with a height of 140~mm.
Once the cell was filled with beads, a load, typically 7600~ N, was applied to the upper piston using a pneumatic press while the lower piston was held fixed.  
In most experimental runs, the outside cylinder wall was not connected to either piston so that the cylinder was supported only by friction with the bead pack (see Fig.~\ref{fig:apparatus}).   
We shall refer to this as the ``floating wall'' method.  
The system could also be prepared with the bottom piston rigidly attached to the cylinder wall, which we shall refer to as the ``fixed wall'' method.  
To estimate the bead-bead and bead-wall static friction coefficients, we glued beads to a plate resting on another glass or acrylic plate and inclined the plates until sliding occurred.  
We found the static coefficient of friction to be close to 0.2 for both glass-glass and glass-acrylic contacts.  

As the beads were loaded into the cell, they naturally tended to order into a 2D polycrystal along the lower piston.  
The beads against the upper piston, by contrast, were irregularly packed.  
We were able to enhance ordering on the lower piston by carefully loading the system, or disturb it by placing irregularly shaped objects against the surface which were later removed.  
For some experiments, the cell was inverted during or after loading with beads.  
By varying the experiment in these ways, we probed the effect of system history on the distribution of forces.

Contact forces were measured using a carbon paper technique \cite{delyon90,dufresne94,liu95}.  
With this method, all constraining surfaces of the system were lined with a layer of carbon paper covering a blank sheet of paper.  
For the blank sheet we used color copier paper, which is smoother, thicker, and has a more uniform appearance than standard copier paper.   
Beads pressed the carbon onto the paper in the contact region and left marks whose darkness and area depended on the force on each bead.  
After the load had been applied to the bead packing, the system was carefully disassembled and the marks on the paper surface were digitized on a flatbed scanner for analysis.  
A region from a typical data set taken from the area over one of the pistons is shown in Fig.~\ref{fig:apparatus}.  
Each experiment yielded approximately 3,800 data points over the interior cylinder wall and between 800 and 1,100 points for each of the piston surfaces, depending on how the system was prepared.  
The position of each mark was identified and the thresholded area and integrated darkness were calculated.  
At the scan resolution used, marks ranged from several pixels to several hundred pixels in area.

\begin{figure}
\centerline{
\psfig{file=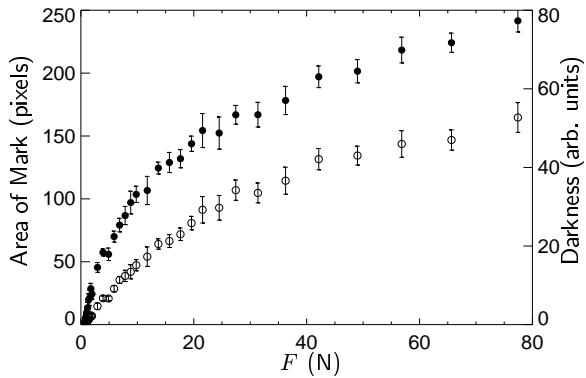,width=3in}}
\vspace{2ex}
\caption{
Calibration curves for the conversion of pressure mark size or intensity to normal force.  
The solid circles represent the mark area and the open circles its integrated darkness.
}
\label{fig:calibration}
\end{figure}

The force was determined by interpolating the measured area and darkness on calibration curves that were obtained by pressing a single bead with a variable, known force onto the carbon paper.  
This was achieved by slowly lowering a known mass through a spring onto a single bead.  
The spring was essential as it greatly reduced the otherwise large impulse which occurs when a bead makes contact with the carbon paper and quickly comes to rest.  
Both area and darkness of the mark left on the copier paper were found to increase monotonically with the normal component of the force exerted by each bead, as seen in Fig.~\ref{fig:calibration}.  
Note that the only requirement is that these curves are monotonic; we do not assume any particular functional relationship. 
With this carbon paper technique, we were able to measure forces between 0.8~N and 80 N with an error of less than 15\%.
We ensure that the beads do not slide relative to the carbon paper during an experiment by measuring the eccentricity of each mark.
We find that the eccentricities $\epsilon$ are narrowly distributed with a mean of 0.1, corresponding to a ratio of major to minor axis $ {a \over b} = {1\over{\sqrt{1-{\epsilon^2}}}}$ of 1.005 for both piston surfaces and container walls.

We find that for less than approximately 0.8~N, little or no mark is left on the copier paper.  
A consequence, visible in Fig.~\ref{eq:ourfit}, is that there are regions where there may have been one or more contacts with normal force less than 0.8~N, or alternatively, which may have had no bead in contact with the surface.  
This ambiguity presents a problem for the precise determination of the mean force  $\overline{F}$.  
To estimate the number of contacts below our resolution, we could fill the voids with the maximum possible number of additional beads, using a simple computer routine.  
However, this over-estimates the number of actual contacts with the carbon paper.  
Instead, we used the following method:  
The average number of beads touching a piston surface was measured by placing double-sided tape on the piston and lowering it onto the pack.  
The tape was sufficiently sticky that the weight of a single bead would affix it to the tape.  
Subtracting the average number of contacts with $F>0.8$~N from this number, we found that 6.4\% of the beads on the lower piston and 4.3\% of the beads on the upper piston have $F<0.8$~N.  
The upper piston had fewer points below 0.8~N because the total number of beads in contact with that piston was typically smaller than on the bottom, raising the mean force and decreasing the fraction of beads with $F<0.8$~N.  
The weight supported by the walls was calculated by subtracting the net weight on the two pistons.  
For experiments performed with floating walls, we verified that the pistons had equal net force (since the weight of the walls can be neglected with respect to the applied force).

\section*{Results}
\begin{figure}
\centerline{
\psfig{file=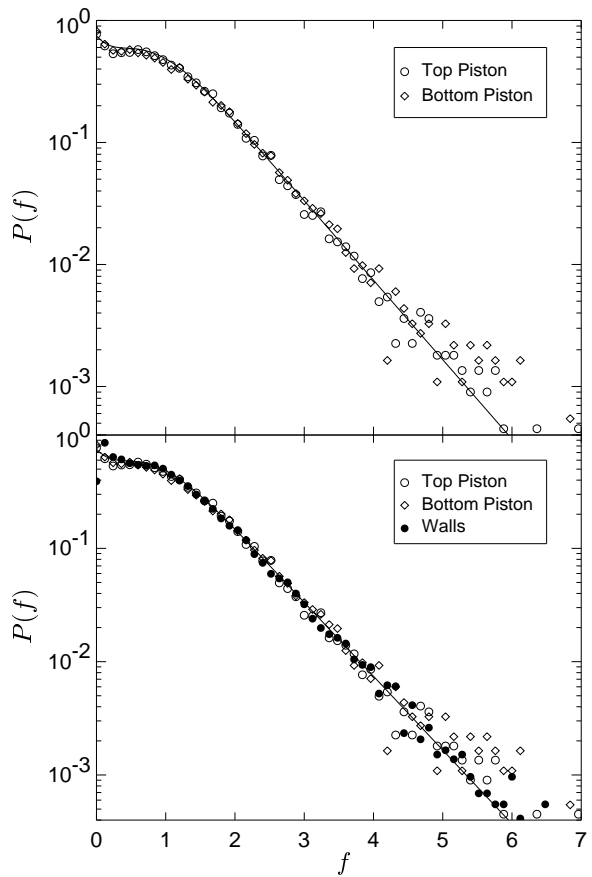,width=3in}}
\vspace{2ex}
\caption{
The distribution $P(f)$ of normalized forces $f$ against the top piston (open circles), the bottom piston (diamonds), and the walls (solid circles).  
The upper panel shows $P(f)$ for the pistons, averaged over fourteen identical experiments.   
The curve drawn is a fitting function as explained in the text (Eq.~\ref{eq:ourfit}).  
The lower panel shows the same data, but with data from the walls included as well.
}
\label{fig:distribution}
\end{figure}

While we conducted experiments with both fixed walls and floating walls, most experiments were performed with the walls floating to reduce asymmetry. 
In this configuration the cylindrical wall of the system was suspended solely by friction with the bead pack.  
Since the applied load was much greater than the weight of the system, any remaining asymmetry between the top and bottom of the system must have come primarily from system preparation, and not from gravity.  
In Fig.~\ref{fig:distribution} we show the resulting force distributions $P(f)$ (where $f \equiv {F/{\overline{F}}}$ is the normalized force) for all system surfaces, averaged over fourteen experimental runs performed under identical, floating wall conditions.  
We find that, within experimental error, the distributions $P(f)$ for the upper and lower piston surfaces are identical and, in fact, independent of floating or fixed wall conditions.  
Note that the lowest bin contains forces from 0~N to roughly 1~N which includes both measured forces as well as an estimated number of undetectable contacts, giving it a greater uncertainty than other bins.  
For forces greater than the mean ($f>1$), the probability of a bead having a certain force decays exponentially,$P(f) \propto e^{-\beta f}$, with $\beta = 1.5 \pm 0.1$.  

Also shown in Fig.~\ref{fig:distribution} is a curve corresponding to the functional form 
\begin{equation}
P(f)=a(1-be^{-f^2})e^{-\beta f}.
\label{eq:ourfit}
\end{equation} 
An excellent fit to the data is obtained for $a$=3, $b$=0.75, and $\beta$= 1.5.  
This functional form captures the exponential tail at large $f$, the flattening out of the distribution near $f\approx 1$, and the slight increase in $P(f)$ as $f$ decreases towards zero.

\begin{figure}
\centerline{
\psfig{file=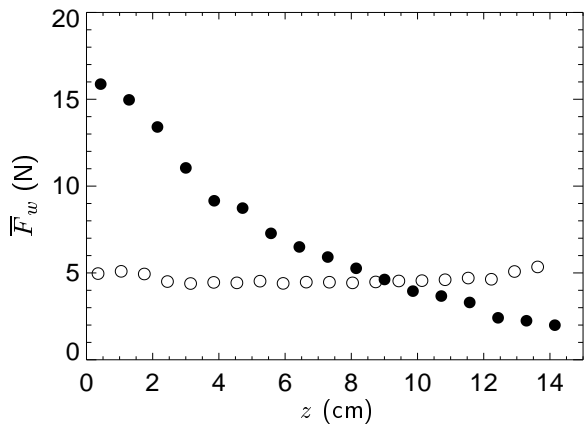,width=3in}}
\vspace{2ex}
\caption{
The mean normal force $\overline{F}_w(z)$, measured along the wall at height $z$ below the top surface of the packing, for fixed wall (solid circles) and floating wall (open circles) boundary conditions.
}
\label{fig:wallforces}
\end{figure}

For the mean force against the side wall we observe a dependence with the depth, $z$, into the pile from the top piston which strongly depends on the boundary conditions (Fig.~\ref{fig:wallforces}).  
For fixed wall boundary conditions (solid symbols) the angle-averaged wall force, $\overline{F}_w(z)$, is greatest near the upper piston, decaying with increasing depth into the pile. 
On the other hand, for floating wall conditions (open symbols) $\overline{F}_w(z)$ stays roughly constant.  
Using $\overline{F}_w(z)$ we compute the set of normalized forces, $f_{w,i} \equiv {F_{w,i}/{\overline{F}_w(z_i)}}$, exerted by individual beads, $i$, against the side walls.  
We find that the probability distribution, $P(f_w)$, is independent of $z$ within our experimental resolution and is practically identical to that found on the upper and lower piston surfaces, with a decay constant $\beta_w = 1.5 \pm 0.2$ for the regime $f_w > 1$.  
This distribution is shown in Fig.~\ref{fig:distribution} by the solid symbols.  
Since along the walls we were unable to determine directly the number of contacts with force less than 0.8~N, we estimated it to be 4.3\%, based on our result for the disordered piston.  
The uncertainty in $\beta_w$ is predominantly due to the uncertainty in this estimate.  
Note that within the resolution of our measurements, the probability distributions in Fig.~\ref{fig:distribution} are the same for all surfaces.

In contrast to observations reported previously \cite{liu95,kuno83}, we observe that the mean force on any portion of the piston is independent of position.  
The radial dependence of the mean force against the pistons found previously \cite{liu95} was an artifact of the compression method, and does not occur if the load is applied using a pneumatic press with carefully aligned pistons.

The first few layers of monodisperse beads coming into contact with the lower piston tend to order in a hexagonal packing while farther into the system a random packing is observed.  
To probe the effect of boundary-induced crystallization, the degree of bead ordering was varied in some experiments.  
We used the measured positions of the marks left on the copier paper to compute the radial distribution function,
\begin{equation}
g(r)= {1 \over{Nn_o\pi r}} \sum_{i=1}^{N} \sum_{j=i+1}^{N} \delta (r_{ij} - r)
\label{eq:gofr}
\end{equation}
where $n_o$ is the average density of points and $r_{ij}$ is the distance between the centers of marks $i$ and $j$.  
If filled from the bottom up without container inversion, the packing structure over the lower piston surface clearly exhibits a larger degree of crystalline order than that touching the top piston surface, as seen in Fig.~5a,b.  
Vertical lines are drawn to indicate peaks expected in $g(r)$ for a 2D hexagonal packing.  
The radial distribution function for the lower piston in an experiment where ordering along this piston is disturbed is shown in Fig.~5c.  
Despite the significant differences in degree of ordering evident from Fig.~5a-c, no significant effect on $P(f)$ was observed.

\begin{figure}[t]
\centerline{
\psfig{file=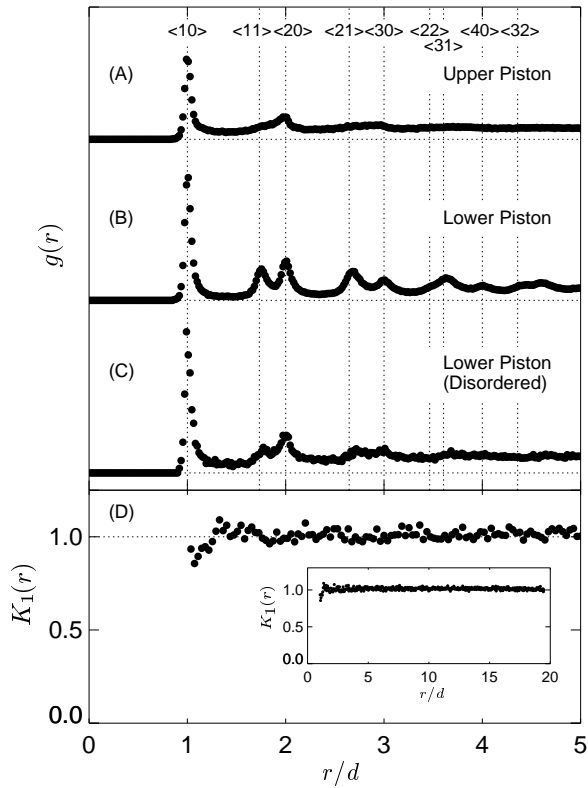,width=3in}}
\vspace{2ex}
\caption{
Pair distribution function $g(r)$ for (a) upper piston, (b) lower piston, and (c) lower piston with disrupted ordering. 
The horizontal axis gives the distance, $r$, between any two points, normalized by the bead diameter $d$.
Vertical lines indicate the distances between points  separated by hexagonal lattice translation vectors and are labeled by the vector indices.   
(d) Force pair correlation function $K_1(r)$ for the bottom piston.
The inset shows $K_1(r)$ out to 20 bead diameters, a distance equal to the radius of the cell and half its height.
}
\label{fig:correlations}
\end{figure}

Since beads generally move downward as the cell is loaded, friction forces tend to be oriented upward.  
The process of adding beads to fill the cell, therefore, breaks the symmetry of the system by building an overall directionality into the force network.  
With different packing histories, however, such as inverting the system once or more during or after loading, we systematically disrupted this directionality.  
Again no measurable effect on $P(f)$ was found. 

Our experiments also allowed for a direct calculation of correlations between normal forces impinging on a given container surface.  
We computed the lateral force-force pair correlations
\begin{equation}
K_n(r)= {\sum_{i=1}^N \sum_{j=i+1}^N \delta (r_{ij}-r) f_i^n f_j^n \over  \sum_{i=1}^N \sum_{j=i+1}^N \delta (r_{ij} -r) }
\label{eq:kofr}
\end{equation}
over both piston surfaces and the walls.  
As an example, Fig.~5d shows the first order correlation, $K_1(r)$, for the lower piston in experiments where ordering was not disrupted (corresponding to $g(r)$ in Fig.~5b).  
The featureless shape of $K_1(r)$ is characteristic of all cases examined ($n \in$ \{1,2,3\}) and indicates no evidence for force correlations.

\section*{Discussion}
The key features of the data in Fig.~\ref{fig:distribution} are the nearly constant value of the probability distribution for $f<1$ and the exponential decay of $P(f)$ for larger forces.  
No comprehensive theory exists at present that would predict this overall shape for $P(f)$.  
The exponential decay for forces above the mean is predicted by the scalar q-model as a consequence of a force randomization throughout the packing \cite{liu95,coppersmith96}.  
In this mean field model the net weight on a given particle is divided randomly between $N$ nearest neighbors below it, each of which carries a fraction of the load.  
Only one scalar quantity is conserved, namely the sum of all force components along the vertical axis. 
Randomization has an effect analogous to the role played by collisions in an ideal gas \cite{liu95,coppersmith96}.  
The result is a strictly exponential distribution $P(f) \propto e^{-Nf}$ for the normal forces across the contact between any two beads.   

The calculations for the original q-model were done for an infinite system without walls.  
If one assumes that each particle at a container boundary has $N$ neighbors in the bulk and a single contact with the wall, then the net force transmitted against the wall is a superposition of $N$ independent contact forces on each bead, so that the probability distribution for the net wall force is modified by a prefactor $f^{N-1}$, much in the way a phase-space argument gives rise to the power law prefactor in the Maxwell-Boltzmann distribution.  
Thus, the original q-model predicts a non-monotonic behavior for $P(f)$ with vanishing probability as $f \rightarrow 0$.  
Such a ``dip'' at small force values has also been found in recent simulations by Eloy and Clement \cite{eloy97}.  
It is, however, in contrast to the data in Fig.~\ref{fig:distribution} and to recent simulation results on 2D and 3D random packings by Radjai and coworkers \cite{radjai96b,radjai97c,radjai97b}.  
These simulations indicated that the distribution of normal contact forces anywhere, and at any orientation, in the packing did not differ from that found for the subset of beads along the walls.   
In fact, for both normal and tangential contact forces inside and along the surfaces of the packings, Radjai et al. observed distributions that were well-described by  
\begin{equation}
P(f) \propto 
\cases{ f^{-\alpha} & $f<1$\cr
e^{-\beta f} & $f>1$\cr}
\label{eq:radjaipf}
\end{equation}
with $\alpha$ close to zero and positive and $1.0<\beta<1.9$, depending on which quantity was being computed, the dimension of the system, and the friction coefficient.  
While we were unable to experimentally measure forces below about $f \approx 0.1$,  the simulation data by Radjai and coworkers extends to $f \approx 0.0001$.
Power law behavior with $\alpha > 0$ in Eq.~\ref{eq:radjaipf}, if indeed correct, would lead to a divergence in $P(f)$ as $f \rightarrow 0$.  
However, we observe that our empirical function, Eq.~\ref{eq:ourfit}, which does not diverge, provides a fit essentially indistinguishable from a power law $f^{-\alpha}$ over the range  $0.001<f<1$ as long as $\alpha$ is positive and close to zero.  
We can thus equally well fit the simulation data for normal forces in Refs. \cite{radjai96b,radjai97c,radjai97b}, over its full range, with Eq.~\ref{eq:ourfit}.  
For the case of 3D simulations and friction coefficients close to 0.2, this is possible using the same coefficients as for the experimental data in Fig.~\ref{eq:kofr}. 

We point out that the fitting function in Eq.~\ref{eq:ourfit} is purely empirical.  
In particular, we do not have a model that would predict the $(1-be^{-f^2})$ prefactor of the main exponential.  
It may be possible to think of this prefactor, in some type of modified q-model, as arising from considerations similar to phase-space arguments.  
The fact that it clearly differs from the usual $f^N$ dependence expected for $N$ independent vector components would then point to the existence of correlations between the contact forces on each bead.  
Such correlations obviously exist, in the form of constraints; yet how these constraints conspire to give rise to a specific functional form for $P(f)$ as in Eq.~\ref{eq:ourfit} remains unclear.  
Eloy and Clement \cite{eloy97} have attempted to take into account some of the correlations that might apply to forces acting locally on a given bead.  
Using a modified q-model they include the possibility of a bias in the distribution of $q$'s, leading to a screening of small contact forces by larger ones.  
The resulting $P(f)$, nevertheless, still tends to zero as $f \rightarrow 0$.

Finally, we note that a ``dip'' in $P(f)$ for small forces can always be introduced by averaging our data over areas large enough to contain several pressure marks.  
Data by Miller et al. \cite{miller96} on shear cells, using stress transducers of various sizes, similarly show an increasingly pronounced ``dip'' for the larger transducers.  
They did not, however, observe the pronounced narrowing of the distribution that is expected in the limit of sufficiently large areas and attributed this to possible force correlations.  
Our data for the force pair correlations in Fig.~\ref{fig:correlations} indicate that no simple correlations exist between forces within the plane of any of the confining walls.  
This result is in accordance with the q-model \cite{narayanpc}.

\section*{Conclusion}
We have found that the distribution of forces, shown in Fig.~\ref{fig:distribution}, is a robust property of static granular media under uniaxial compression.  
Its shape turns out to be identical, within experimental uncertainties, for all interior container surfaces and furthermore appears to be unaffected by changes in the boundary conditions or in the preparation history of the system.  
The exponential decay for forces above the mean emerges as a key characteristic of the force distribution.  
The exponential tail of the distribution can be understood on the basis of a scalar model (q-model), where it emerges as a result of a randomization process that occurs as forces are transmitted through the bulk of the bead pack.  
The consequences of the vector nature of the contact forces on the distribution, however, remain unclear.  
A second key aspect of the measured distribution is the absence of either a ``dip'' or a powerlaw divergence for small forces; instead, our data is most consistently fit by a functional form that approaches a finite value as $f \rightarrow 0$.  
This empirical fitting form, Eq.~\ref{eq:ourfit}, provides an excellent fit over the full range of forces for our experimental data, as well as for simulation results on 3D packings obtained by Radjai et al. and for simulations performed by Thornton.

\section*{Acknowledgments}
We would like to thank Sue Coppersmith, John Crocker, David Grier, Hans Herrmann, Chu-heng Liu, Onuttom Narayan, Farhang Radjai, David Shecter, and Tom Witten for many useful discussions.  
This work was supported by the NSF under Award CTS-9710991 and by the MRSEC Program of the NSF under Award DMR-9400379.

\end{document}